# WANDERLUST: 3D IMPRESSIONISM IN HUMAN JOURNEYS

**Guangyu Du**
Senseable City Laboratory
Massachusetts Institute
of Technology
dugy.guangyu@gmail.com

**Fábio Duarte**
Senseable City Laboratory
Massachusetts Institute
of Technology
fduarte@mit.edu

**Lei Dong**
Senseable City Laboratory
Massachusetts Institute
of Technology
Peking University, China
arch.dongl@gmail.com

**Carlo Ratti**
Senseable City Laboratory
Massachusetts Institute
of Technology
ratti@mit.edu

**Abstract**
The movements of individuals are fundamental to building and maintaining social connections. This pictorial presents Wanderlust, an experimental three-dimensional (3D) data visualization on the universal visitation pattern revealed from large-scale mobile phone tracking data. It explores ways of visualizing recurrent flows and the attractive places they implied. Inspired by the 19th-century art movement Impressionism, we develop a multi-layered effect, an 'impression', of mountains emerging from consolidated flows, to capture the essence of human journeys and urban spatial structure.

**Introduction**
Today, more than half the world's population lives in cities [14]. To understand how individuals from different urban areas share the same space, forge and maintain social connections, it is crucial that we know where people go and how often they do so. However, previous models of population-level flows, such as the gravity law and the radiation model, mainly focus on spatial dependence, thereby being insufficient to reveal the heterogeneity of urban trips that span temporal scales [1].

The universal visitation law of human mobility [12], published recently in *Nature*, identified the missing component by depicting the interplay between the spatial and temporal spectrum of human journeys. The research finds a robust law for modeling the number of visitors to any location, stating that it decreases as the inverse square of the product of the visiting frequency and travel distance. Moreover, the magnitude of the flows is determined by a location-specific constant, 'attractiveness', which reflects the commonly shared interests towards a place based on its characteristics.

While the elegant distance-frequency law seems slightly abstract for general audiences, the spatial distribution of the location-dependent 'attractiveness' is representational. As urban dwellers, we seldom know the specific number of visitors before gaining a feeling that some places are more attractive than others, for the location, physical layout and functional features, as well as the atmosphere created by strangers.

Our visualization aims to draw on this representational location-specific attractiveness to construct a general impression of how people are connected by places as well as present details on how the distance-frequency law manifests in the recurrent population flows. By shaping flows into mountains, we indicate a reciprocal effect between us and the physical environment we inhabit. Our decisions on journeys shape the 'attractiveness' landscape, which in turn influences our desire to wander — higher mountains are more vivid in our eyes. Through movements from place to place and day to day, we unconsciously become the background for each other's life

In the era of big data with increasingly fine spatial and temporal scales, the design process of this visualization is also a quest for 3D representation as opposed to 2.5D and overall visual effects instead of details. We call this approach 3D Impressionism.

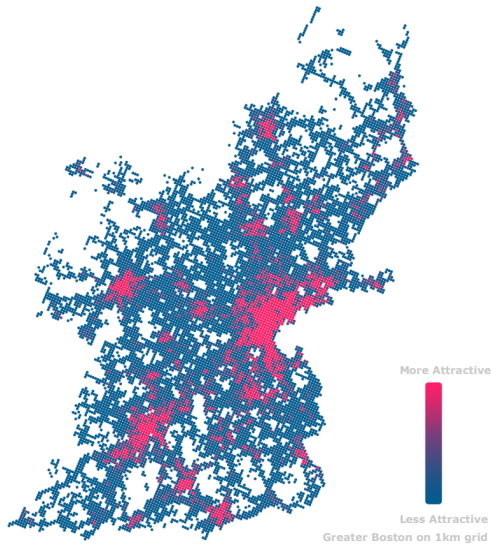

More Attractive

Less Attractive
Greater Boston on 1km grid

**Study Area and Dataset**
The empirical analysis of distance-frequency law is based on five mobile phone datasets: Greater Boston Area in the United States (North America), Singapore (Asia), Lisbon, Porto and Braga in Portugal (Europe), Dakar in Senegal (Africa), and Abidjan in Ivory Coast (Africa). The Greater Boston dataset contains approximately one billion location records from two million anonymized mobile phone users during four months (July to October) in 2009. Since the visitation pattern holds for different months, we visualize only one month (August) for simplicity. The study area is partitioned into a square grid with cells of size 1km by 1km, a resolution that balances the mobile phone tower precisions in metropolitan (~0.5km) and rural (1~2km) areas. Location records are then grouped into these cells.

**Drawing Lines**
To demonstrate the underlying visitation law, for each cell, we want to visualize the direction, magnitude, and frequency of flows coming into it. The challenge would be how to construct a high-level pattern without oversimplifying these details. This has already been a question for most visual explorations of origin-destination (OD) flows when large and complex datasets are increasingly common. For a small number of flows, mapping geographic vectors as lines is straightforward — flow maps have traditionally been used as a cartographic method to portray movements[3]. In the case of big data, such direct flow mapping can suffer from occlusion. And several techniques have been developed to reduce visual clutter of flow maps while retaining the spatial layout:

Force-directed edge bundling combines proximal flows regardless of direction and magnitude [8]. As a generalization of hierarchical edge bundling [6], this algorithm is likely to introduce a tree structure that may not exist in the data [16], making the state of art visualizations similar to each other [10]. Flow density surface rasterizes line density into an image [11]. The resultant surface successfully indicates the spatial structure and the overall interaction patterns between places, yet it detaches from the line layer, thus losing the heterogeneity of flows. OD maps use cells in a two-level gridded spatial treemap to record trajectory density [15]. Inherited from the form of OD matrix, small multiples of geographical space make clusters and associations apparent, but the number of grids supported by this representation is limited.

While these approaches vary in the output drawings, in the form of lines, surfaces, or matrices, they all present mainly the density of links. Thereby the magnitude, direction and frequency of flows remain underdeveloped. Among the three methods, flow density maps and OD maps do provide visual indications for flow direction, though with the help of plotting additional lines or swapping between origin and destination spaces. Built upon force-directed edge bundling, divided edge bundling [13] tries to address these shortcomings by separating edges in antiparallel directions and using blue-to-red color to specify flow direction and width to encode flow magnitude. Here, the design choice of color is a compromised solution, the more effective directional encoding, tapered edges [7], may produce undesired overlapping of edges of varied width when combined with edge bundling. Moreover, occupying the color channel for a spatially dependent variable makes it even more difficult for including additional flow attributes (such as frequency).

## From 2D, 2.5D to 3D

As data records grow to millions and billions, direct depiction is often replaced by summarization and pattern extraction [2]. Instead of looking for occlusion workarounds in two-dimensional (2D) space and worrying about the faithfulness of aggregation, what if we just use the basic direct flow mapping, with the dimensions of visualization space expanded? After all, geographical space is usually 2D after projection, leaving the third dimension intact.

To clarify, encoding information in the third dimension is not guaranteed by a three-dimensional (3D) visualization. A visualization may become 3D simply because it consists of 3D data points that form a non-developable surface. For instance, 3D edge bundling [4] is not conceptually different from 2D force-directed edge bundling. Each link still represents an association between two nodes. Once we have a method to project the decisive 3D object, cortical surface, into lower dimensions, then the whole visualization would not be 3D anymore. Inversely, a representation that uses the third space to encode only derivatives does not fully occupy 3D space. For the widely used 3D arcs

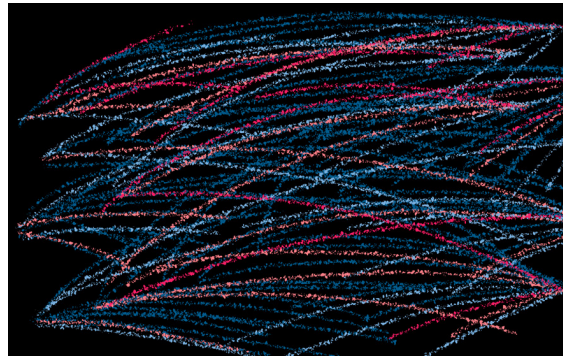

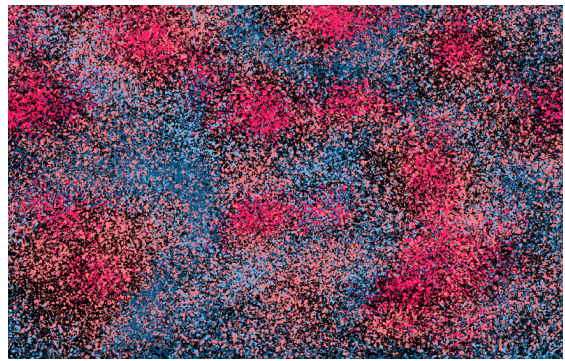

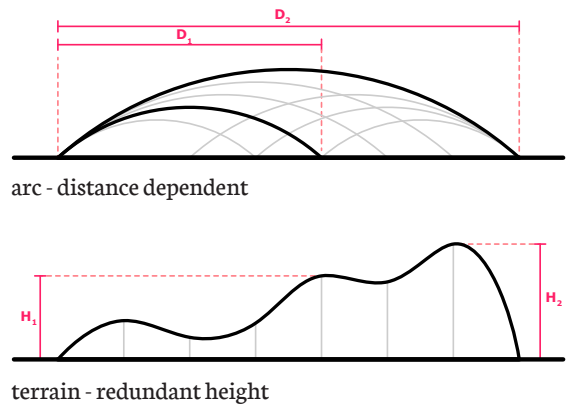

arc - distance dependent

terrain - redundant height

drawn from origins to destinations, layered in different altitudes without overlapping [5], their radius and highest point are dependent on distance. For a 3D terrain that maps the spatial distribution of a variable in its fluctuations of elevations, information encoded by the elevation of vertices could be flattened in 2D through colors or shapes. This also applies to various extruded geometries used in standardized geospatial visualizations.

We propose a 3D solution that combines location-dependent attractiveness and flows. By plotting flows from the origin point on the ground to the destination point at an elevated plane decided by its attractiveness, directed flows between two locations will not interfere with each other, instead, they go into mountains centered at their destinations respectively. This representation has several advantages. First, the previously mentioned flow direction can be easily indicated by a vector from the ground point to the elevated point. Second, color is reserved for encoding additional attributes. Third, overlapping is reduced as collinear lines with different lengths are now separated. Fourth, lines are not distorted.

For specific visual encodings, we choose line width for flow magnitude and color for flow frequency: 1) Line width is commonly used for representing the flow volume in flow maps [9]; 2) the flow frequency in the data only spans 1 to 30 days, easily to be transferred into categorical variables and mapped to colors. 3) There may be cases when hue and brightness are used to map the density of links, such as force-directed edge bundling [8], where all links are drawn in the same width. But this is pixel-level operations - assigning colors by the number of accumulated edges passing through each pixel, not the geometry level - scaling the line widths according to the number of unique visitors (grouped by frequency) in each flow.

Each of the two directions of the flow between A and B contains sub-flows of different frequencies.

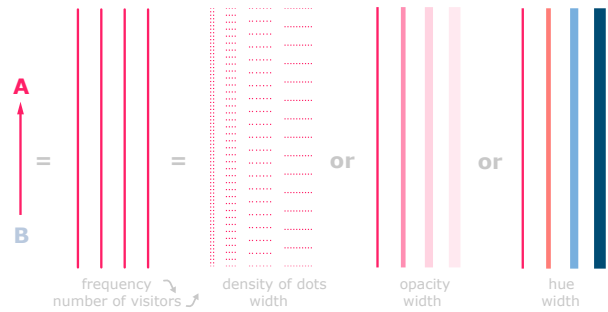

frequency
number of visitors

density of dots
width

opacity
width

hue
width

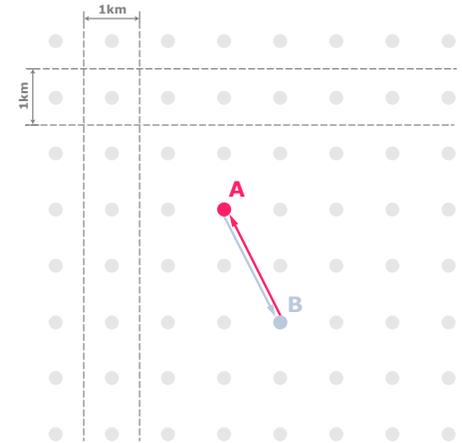

flows to location A will be drawn from the plane at z = 0 (the ground) to the plane at z = μ (attractiveness)

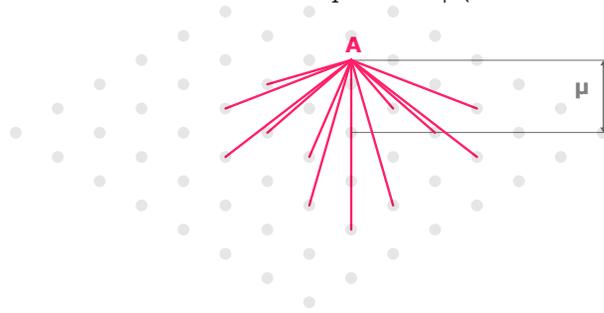

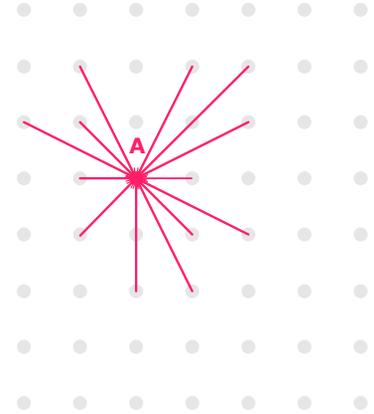

flow from B to A and flow from A to B will not interfere with each other, they contribute to the height of A and B respectively

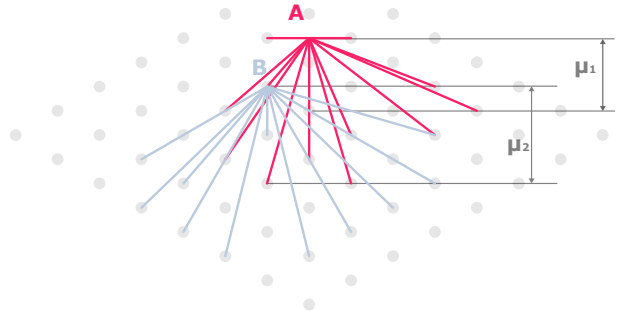

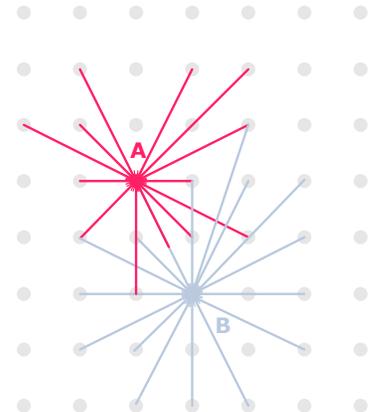

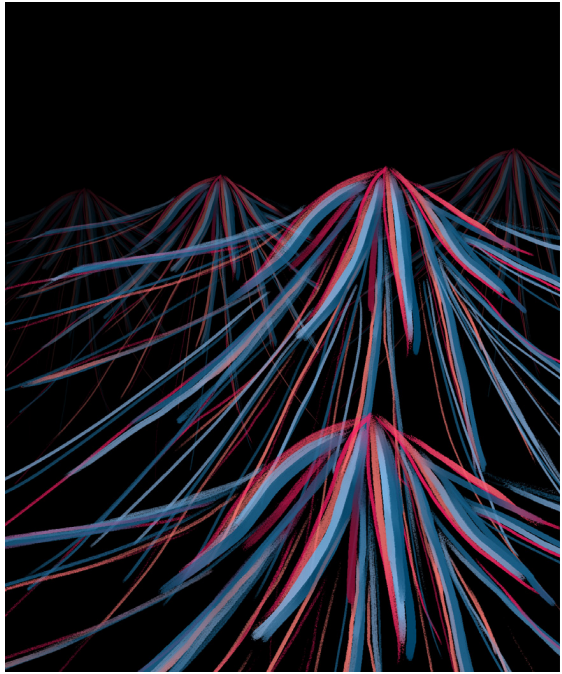

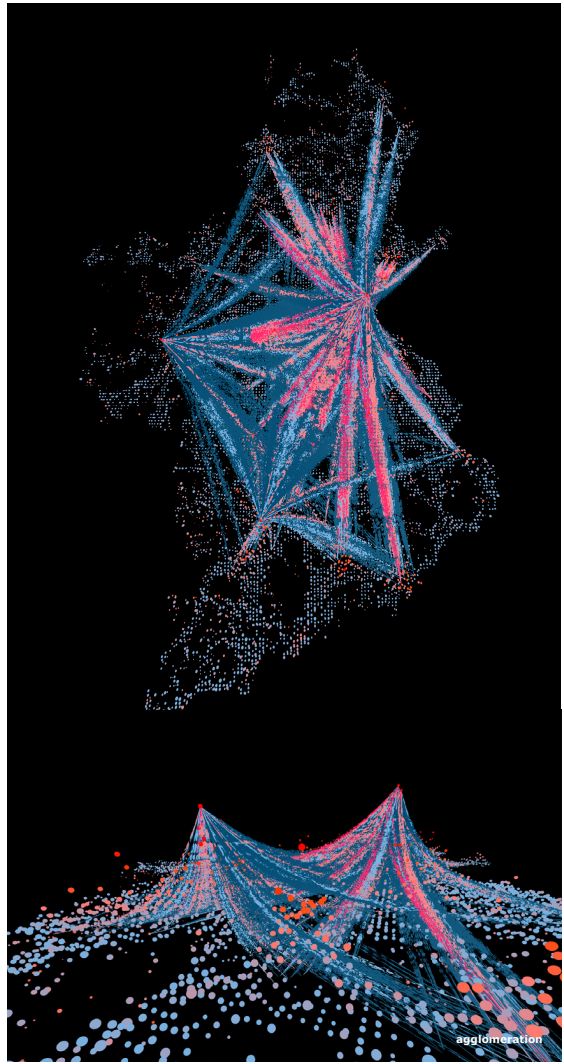

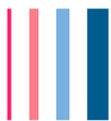
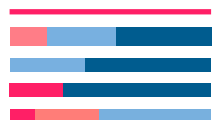
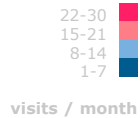

**agglomeration**  **subdivision**  **visits / month**

22-30
15-21
8-14
1-7

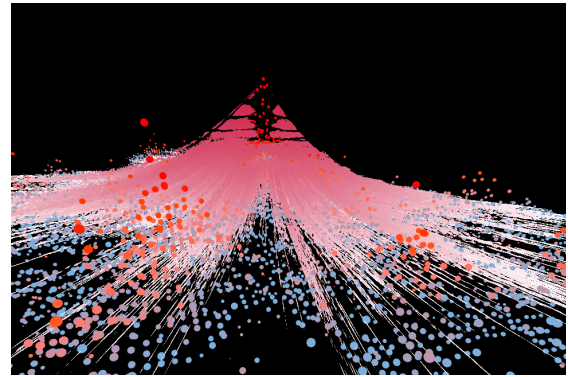

### A Touch of Color

Our initial implementation of the above-mentioned idea is splitting flows into lines colored by frequency groups, with line width decided by the number of visitors in this group. For an OD pair, we draw at most four lines - dark blue represents the visits made less often in the study month, deep pink denotes those frequent journeys, whereas light blue and coral show intermediate movements. This implementation has a few inconveniences. Even though lines are textured to be partially transparent, drawing multiple lines for the same OD pair challenges the depth sorting of pixels. Indeed, each OD pair only needs to visualize once for showing connectivity. Inspired by the color transition used in directed edges [7], we started to think about dividing a flow into color bands along its direction to avoid overdraw.

### From Agglomeration to Subdivision

Going one step further, high spatio-temporal data resolution and visual complexity are not the root causes of the failure brought by juxtaposing lines. The culprit is our need for this visualization to be legible across multiple scales. We hope to develop a scheme that can portray connections within neighborhoods or between towns, depict movements that span weeks or even years. In this case, the agglomeration way - assembling objects for the increasing amount of categorical values or discrete quantitative values is not flexible. On the contrary, the subdivision way - dividing an object based on the relationship of its parts is feasible as such division is infinite within our perceived time and space. Thus, we partitioned each flow into line segments that correspond to different frequency groups. The line width is the total number of visitors in this flow; the length ratio of line segments matches the percentage of individuals that belong to those groups.

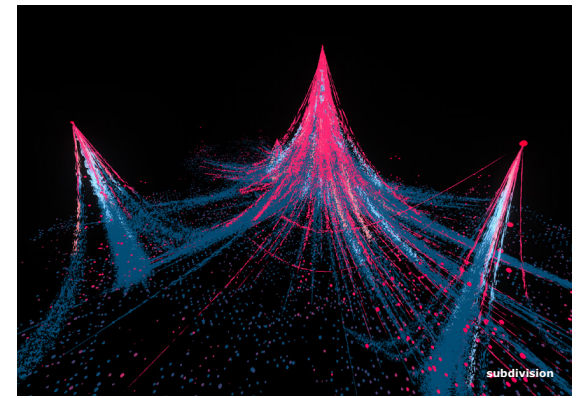

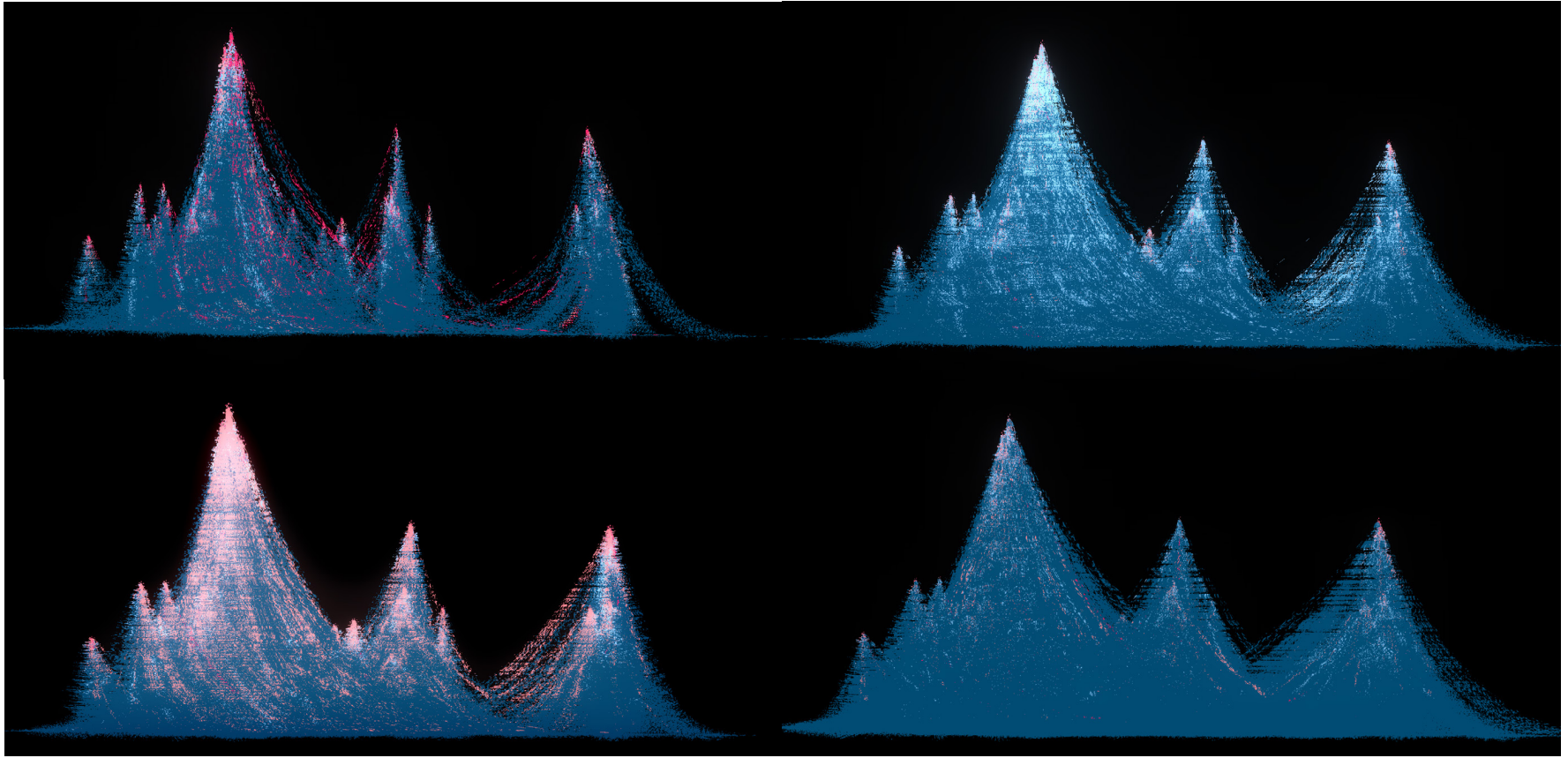

**Towards 'en plein air'**

When inexpensively produced data is becoming one of the principal means through which people perceive the world, the onset of the sensual appeal of data visualization is not far. For web graphics, if we compare scalable vector graphics (SVG) as the studio that trains Academic's perfect lines and contours, then WebGL could be the *en plein air* (outdoor) that stimulates Impressionism's freely brushed colors. Even if the former may scale up to larger data volumes in the future, drawing every single element in vectors makes its visual output more of a meticulous drawing instead of a spontaneous painting, compared with the latter, which only cares about the final rasterized pixels and allows various image processing techniques.

The above four images are produced using WebGL, showing flows with the presence of visitors travelling in four frequency groups. From top-left, bottom-left, top-right to bottom-right, as frequency decreases, we have a feeling of the increased connections and travel distance, corresponding to the distance-frequency law in a qualitative sense.

**Peeling Layers**

In addition to the proliferation of granular data, another trend accompanied by the increased penetration of mobile devices is the emergence of complex systems related to human dynamics. Our stays and movements within and around cities, strangers who tag the places we have just left and friends who receive our messages far away, chatters we have in restaurants and stories we post to Instagram, all become parts of a shared atmosphere captured by ubiquitous location-based services. Our visualization of urban movements and attractive places is featuring only one facet of the above-mentioned 'collective unconscious' or 'sense of togetherness', nurtured by the built environment, technology and media. Yet, it already touches on the characteristics of having layers of complexity.

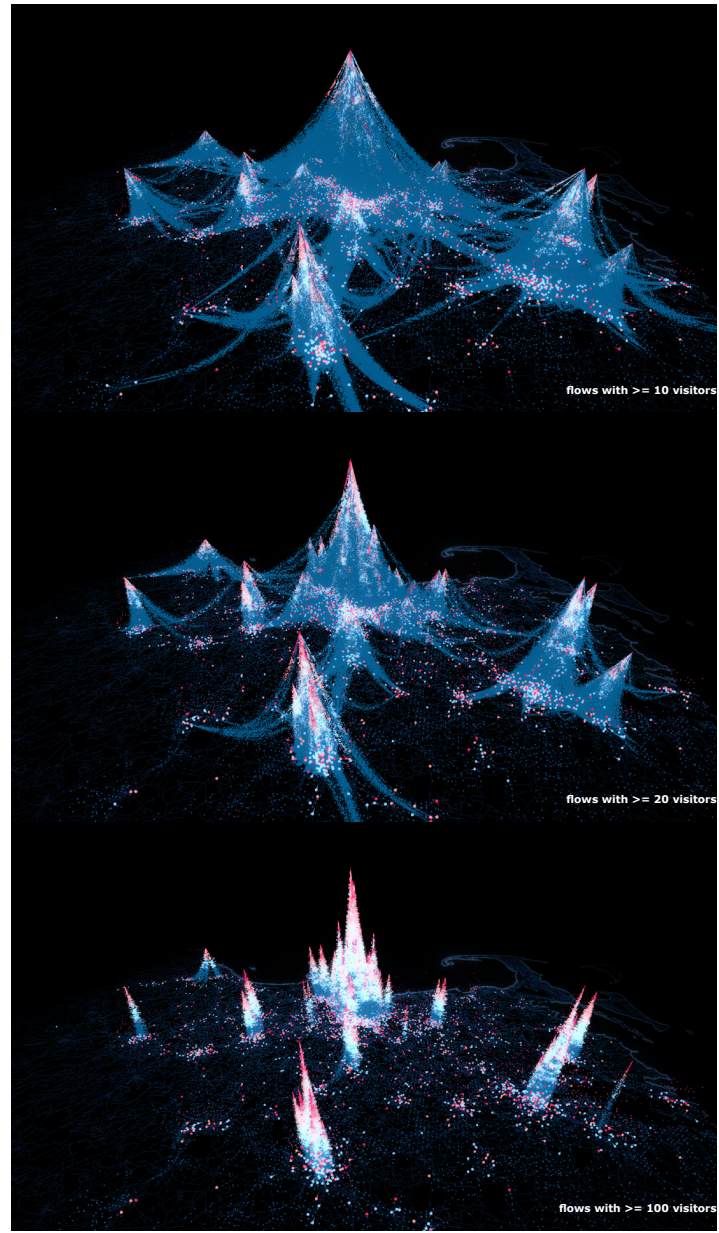

flows with >= 10 visitors

flows with >= 20 visitors

flows with >= 100 visitors

This series of images show flows with unique visitors above a threshold for selected attractive destinations, with unique visitors >=5000 or attractiveness $\log_{10}\mu >= 1.65$.

With looser restriction on visitors count, this image and the image above show more flows than the image below. Flows connecting center areas with the surroundings are mainly movements of infrequent visitors.

This image shows an intermediate layer (visitors >= 50), places near the centers of cities and towns have visitors traveling between low and high frequencies.

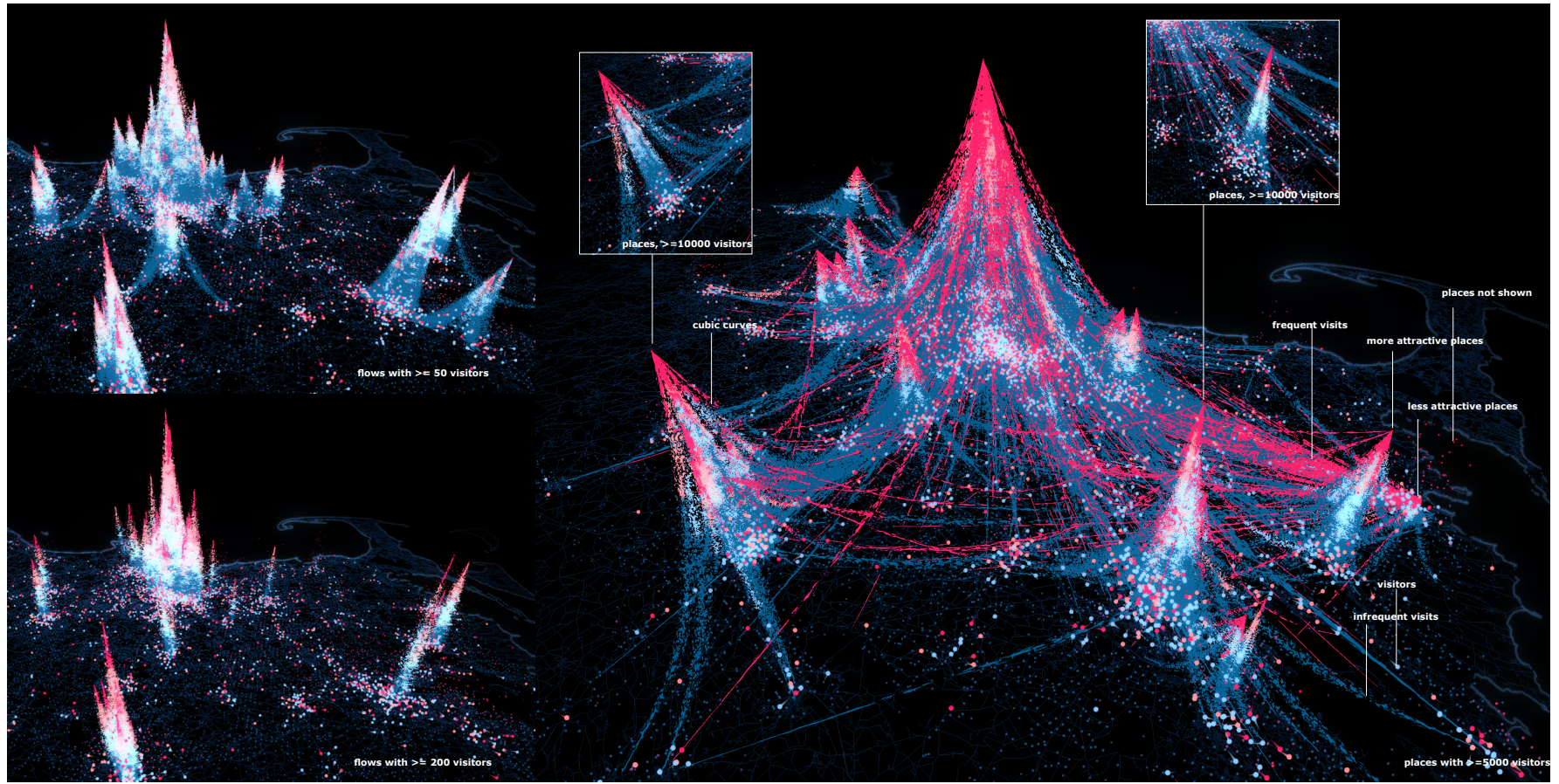

Layers are created by three mechanisms. First, we applied a partially hollowed texture, made with small droplets, to flow lines. It is similar to the use of layered colors in Impressionism — leaving gaps in the top layers to reveal the colors underneath. What viewers see is a mix of colors from flows within a range. Second, we chose cubic curves for flows. For each mountain, peripheral and centralized connections do not interfere with each other, forming layers naturally from the center of a place to its boundaries. Third, we raised[a] the height of mountains to a power based on the location-specific attractiveness.[b] Then in each larger spatial cluster, lower mountains (less attractive places) are covered by higher mountains (more attractive places), making high-level pattern prominent.

Left: Filtering flows with a large amount of visitors, we see they mainly located in the centers of cities and towns, and they often contain frequent visitors.
Right: Compare places with at least 5000 to those of 10000 visitors, smaller mountains appear inside larger mountains that belong to same spatial clusters.

[a] for those non-negative $\log_{10}\mu$ values, which means $\mu > 1$
[b] $\log_{10}\mu$

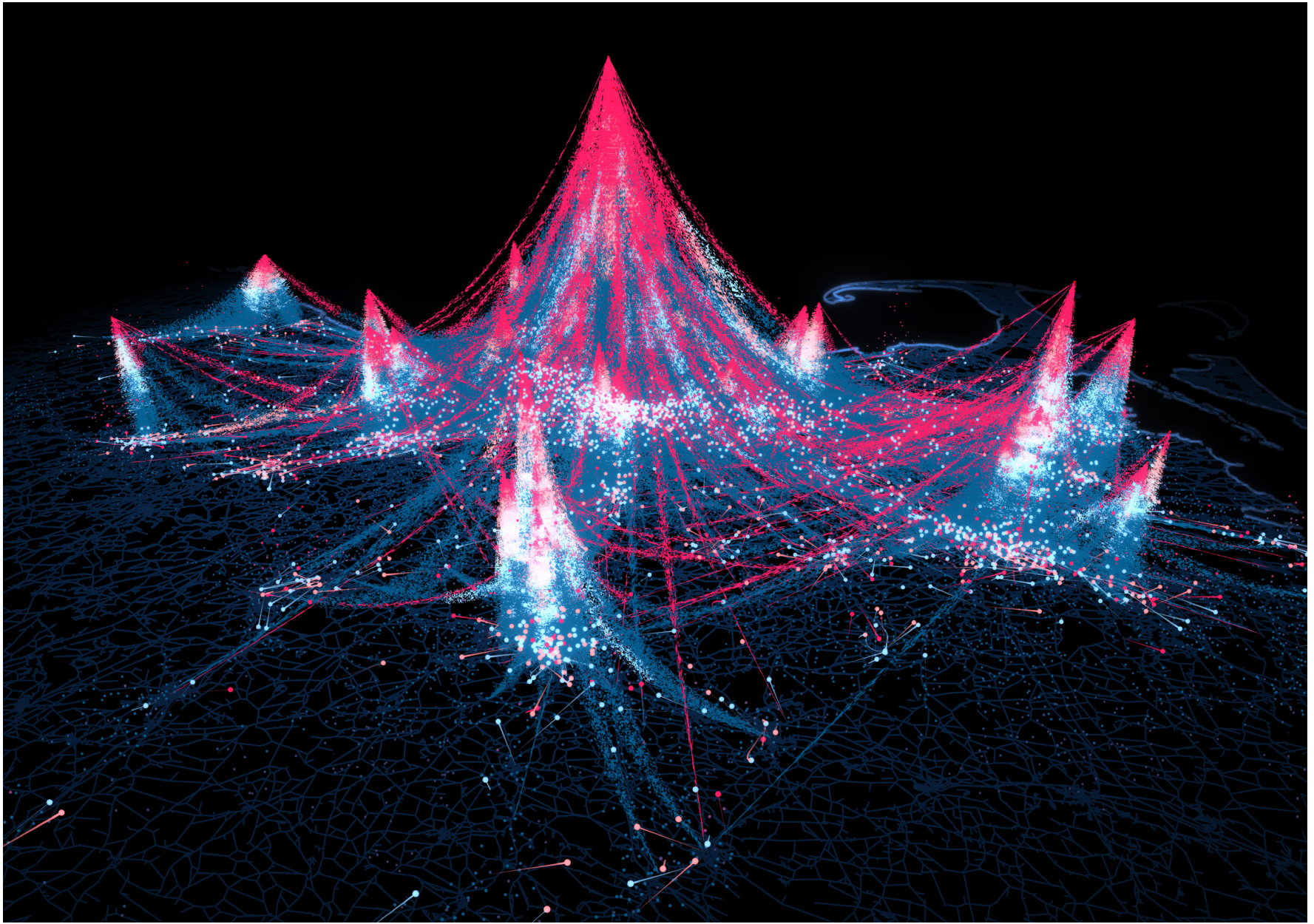

This image visualizes the flows of individuals across the Greater Boston area as lines (visiting frequency as color, number of unique visitors as line width) that form spatial clusters of attractive places, with the height of mountains representing location-specific attractiveness.
destinations: more than 5000 unique visitors within a month or places that have large attractiveness value ($\log_{10}\mu \geq 1.65$)
flows: contain frequent visitors (22-30 visits in a month)

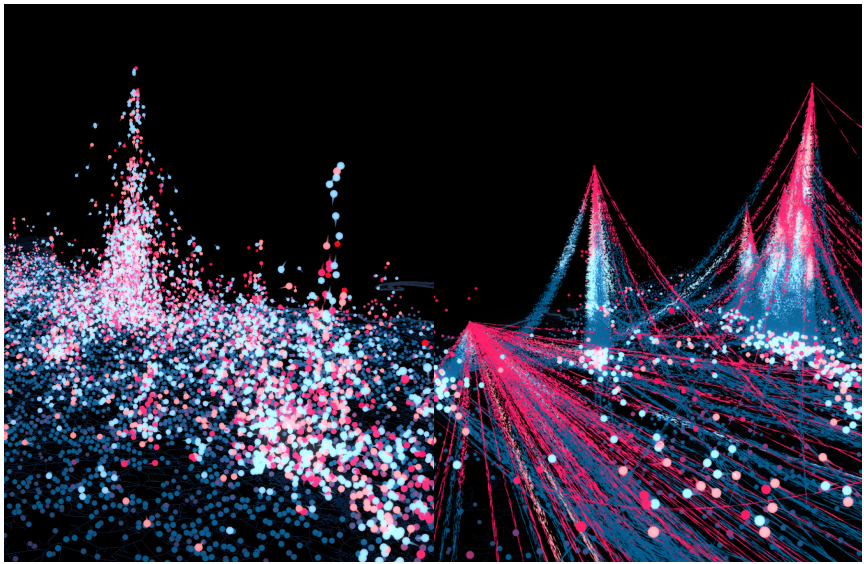

### A Tale of Two Cities

Apart from the above-mentioned three layers that are related to population-level flows, there is another important layer in human journeys — individual behavior. In the original research [12], an individual-level behavior model is built to explain the emergence of spatial clusters. Inspired by it, we want to demonstrate that flows are not staged scenes, but rather the shows that are currently playing in the urban theater. Each flow is a result of one or more individuals traveling occasionally or daily between two places. For performance, we took around 10k (total 340k) visitors to simulate individual movements based on their visitation frequency to previously visited places. At each step, places that have been visited more often are more likely to be chosen as the next destinations. Compared with the optical mixing of colors that requires a lot more points, drawing lines is better at depicting the connections that only have a few individuals traveling with high frequency, as the thin deep pink lines shown in the images above and to the right.

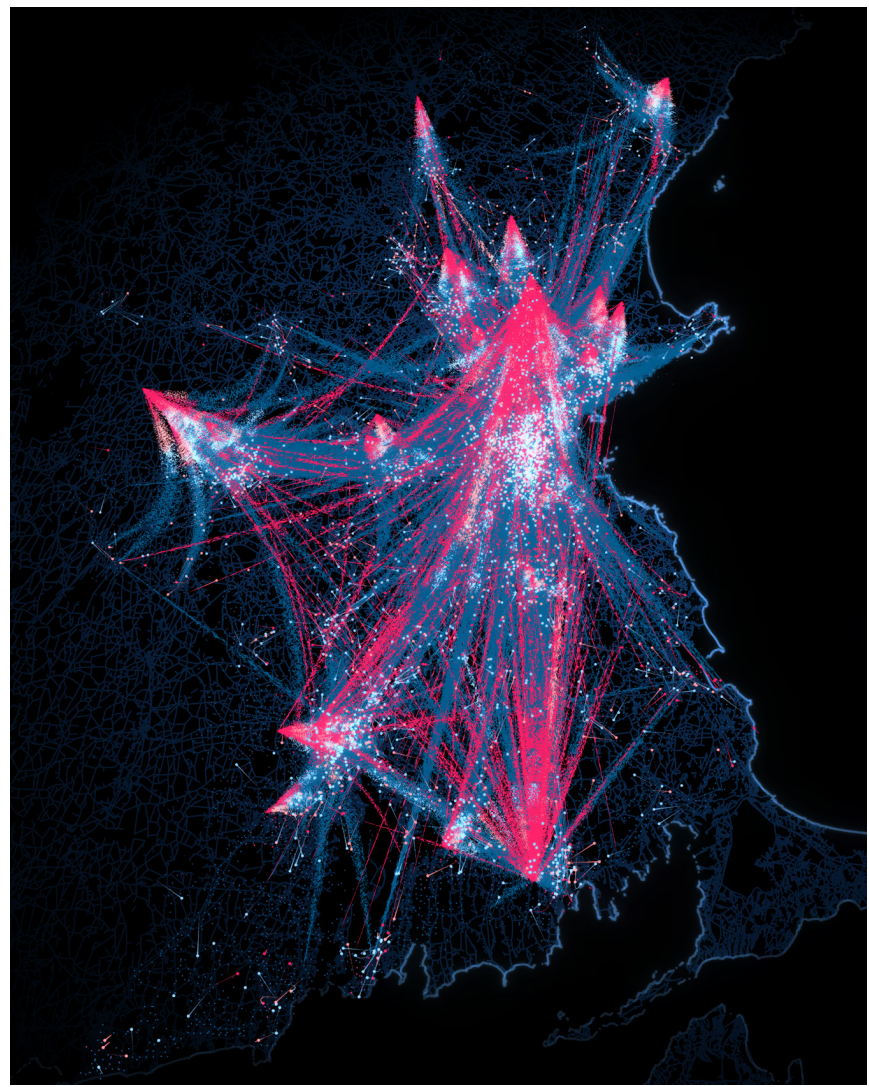

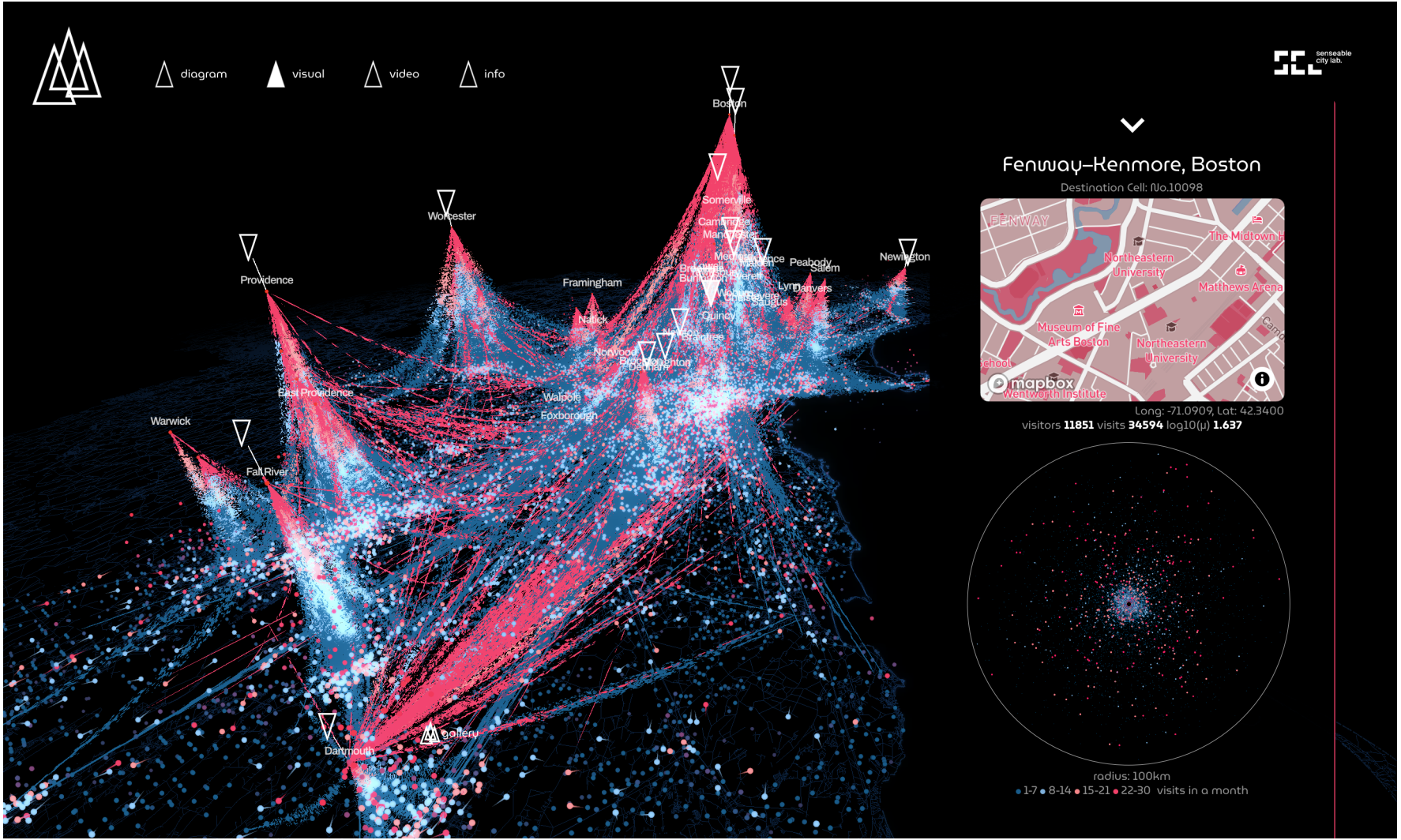

**Wanderlust Interactive Website - Visual Section**

138 attractive destinations are reverse geo-located to 38 attractive towns and 93 attractive neighborhoods( only showing a few selected ones in triangle marks). To the left, flows will disappear, reveal, and change colors; dots that represent visitors will move on mouse dragging. To the right, a panel shows the number of visitors, visits, and the attractiveness, $\log_{10}\mu$, for the current selected cell.

See interactive visuals and diagrams on *https://senseable.mit.edu/wanderlust/*

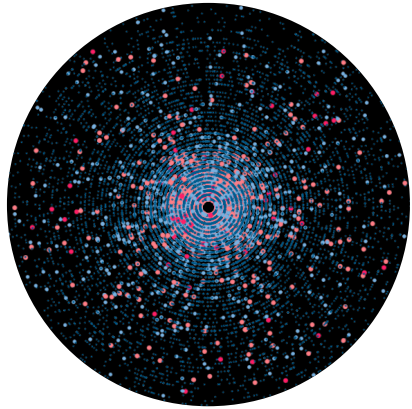

No. 10422  Beacon Hill, Boston

visitors: 29706  visits: 82539  $\log_{10}\mu$: 2.717

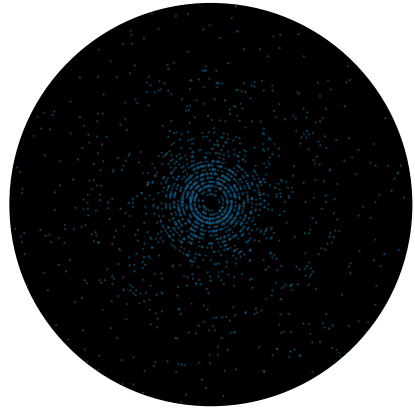

No. 9951  Wright's Park, Medford

visitors: 1951  visits: 2150  $\log_{10}\mu$: 1.919

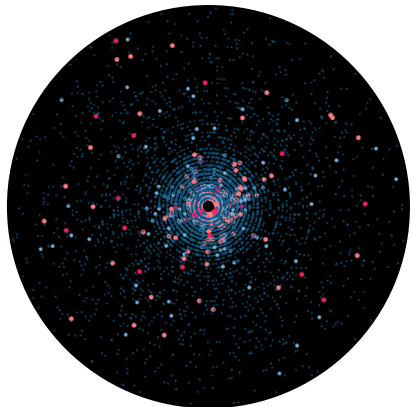

No. 9791  Cambridgeport, Cambridge

visitors: 9339  visits: 20624  $\log_{10}\mu$: 1.705

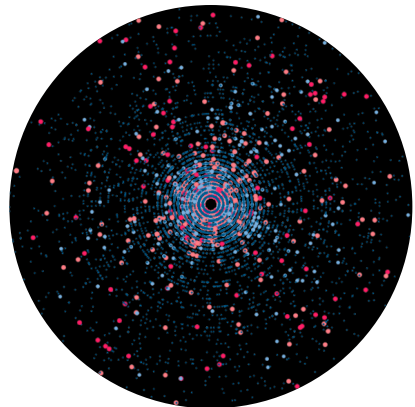

No. 10098  Fenway-Kenmore, Boston

visitors: 11851  visits: 34594  $\log_{10}\mu$: 1.637

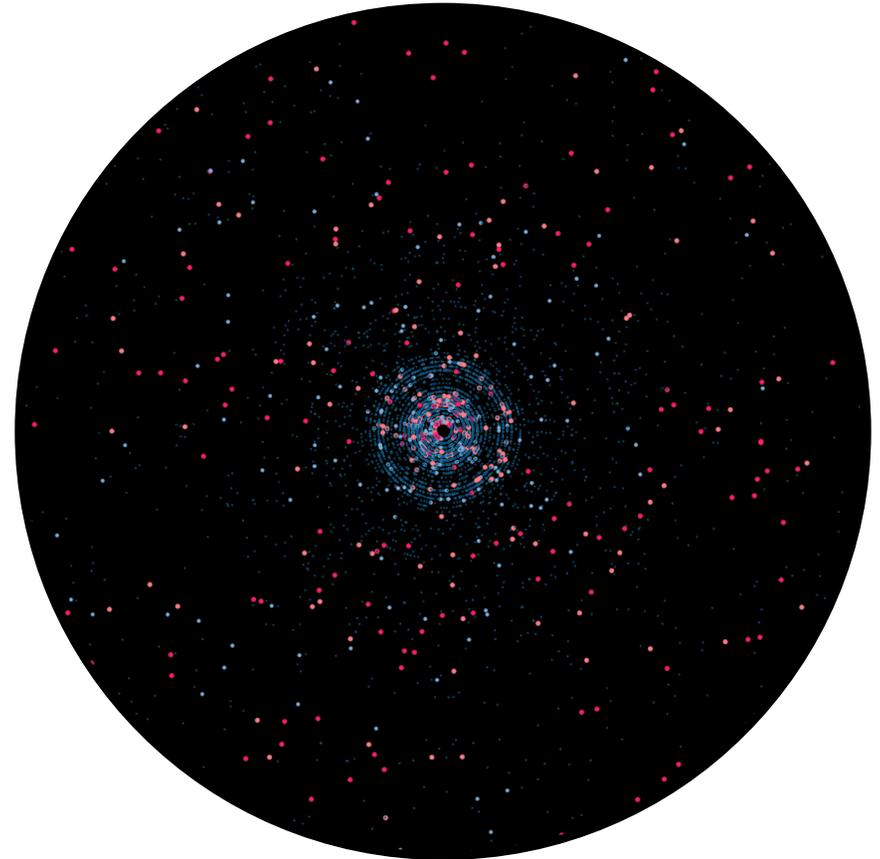

No. 11355  Smith Mills, Dartmouth

visitors: 6180  visits: 19062  $\log_{10}\mu$: 0.919

dots: visitors,  color: frequency,  position: distance to destination

radius: 50km for the left 4 discs,  100km for the one above

**Glimpses of a Moment**

A revision on our previous analogy between SVG's perfect shapes to Academic's art, if the element is small enough, like dots, to function as a small stroke, then the shape itself doesn't influence the freedom of expression. Impressionism sits in between Academic and Expressionism, neither the former's rigor nor the latter's emotion, painting the real world but as if it had been glanced for only a moment.

This series of discs construct snapshots of 'personalities' of attractive places. Places that have similar attractiveness (μ) might differ in the underlying distribution of frequency and travel distance, and the total number of visitors or visits. Though less often, we make great efforts to go to a nature reserve (No. 9951) far away; on the contrary, even if we daily commute to a workplace (No. 9249), it might not be that attractive.

**Discussion- Salon des Refusés**

'Salon of the Refused' is the starting point when artists, including Impressionists, began to arrange their own exhibitions outside the official Paris Salon to draw attention to the new tendencies in art. Today we do not need to worry about the existence of alternative venues — they are all over social media. For us, sovereign individuals empowered by digital networking and e-commerce, the meaning of 'rejection' has changed from lacking exhibitions to the shortage of pigments and data.

To the right, we have applied the proposed new visualization scheme to five urban areas. Despite the open access of two datasets (Senegal and Ivory Coast), the majority of the fine-grained data is not publicly available before research is published. When 19th-century Impressionists have seen the proliferation of new pigments, it is an open question how individuals in our era may be granted access to the inexpensively produced raw data with freedom for exploration but without invasion of privacy. Perhaps with the universal visitation law of human mobility, we can simulate urban dynamics in a mirrored world just as synthesizing cheap cobalt violet.

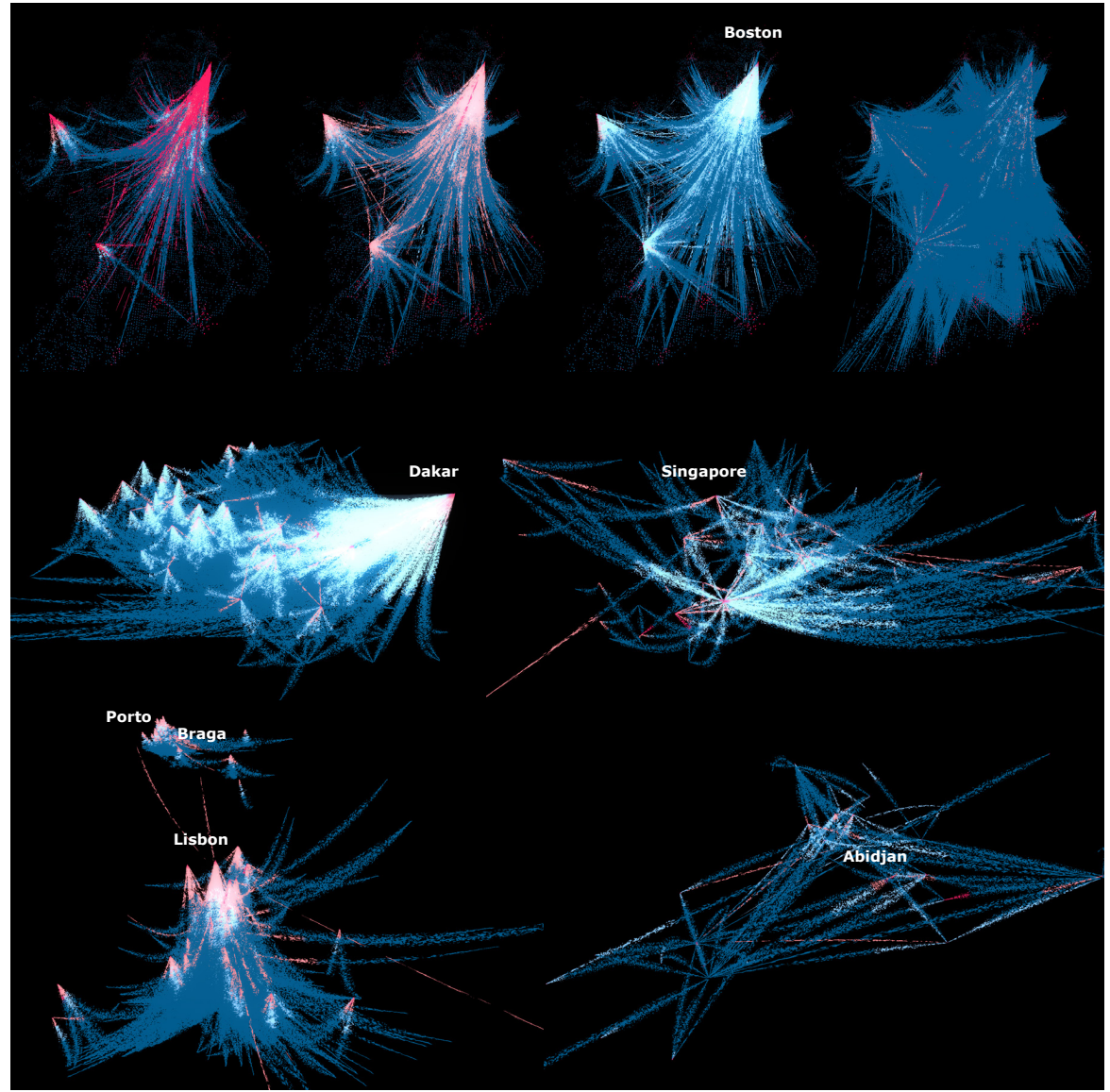

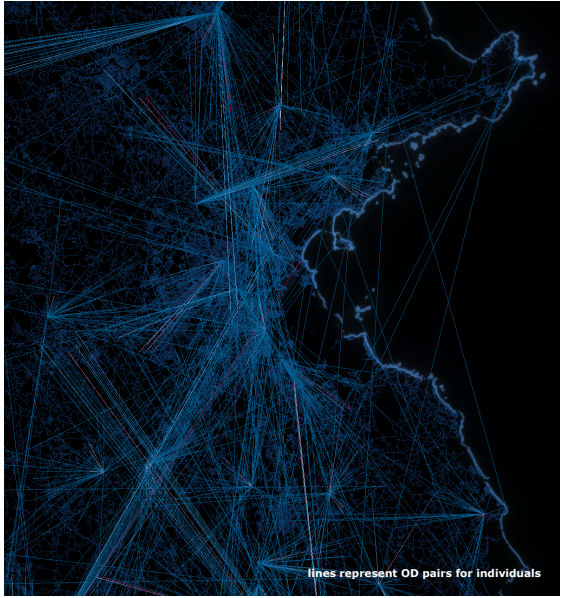

lines represent OD pairs for individuals

For this specific visualization, there are several future directions. First, for frequency groups, if we have finer temporal scales, such as hours or minutes other than current days, interpolated colors might be needed in addition to unmixed colors. Second, it is possible to aggregate current grid-cell-level flows into neighborhood-level or town-level flows with reverse geolocation. Third, current data is fine-grained for population flows but relatively coarse for learning individual behavior. The research assumes people will go back home directly after each visit, but in reality, we often go from place to place during the daytime. With finer spatial scales it is possible to draw similar mountains made by flows for personal trajectories.

VISAP'21, Pictorials and annotated portfolios.

**References**

1. Alessandretti, L., & Lehmann, S. (2021). Trip frequency is key ingredient in new law of human travel. *Nature* 593, 515-516 (2021)
2. Andrienko, G., Andrienko, N., Dykes, J., Fabrikant, S. I., & Wachowicz, M. (2008). Geovisualization of dynamics, movement and change: key issues and developing approaches in visualization research. *Information Visualization* 7 (3–4): 173–80.
3. Claudel, M., Nagel, T., & Ratti, C. (2016). From origins to destinations: the past, present and future of visualizing flow maps. *Built Environment*, 42(3), 338-355.
4. Böttger, J., Schäfer, A., Lohmann, G., Villringer, A., & Margulies, D. S. (2013). Three-dimensional mean-shift edge bundling for the visualization of functional connectivity in the brain. *IEEE Transactions on Visualization and Computer Graphics*, 20(3), 471-480.
5. DECK.GL. ArcLayer Example. Retrieved June 10, 2021 from https://deck.gl/gallery/arc-layer.
6. Holten, D. (2006). Hierarchical edge bundles: Visualization of adjacency relations in hierarchical data. *IEEE Transactions on Visualization and Computer Graphics*, 12(5), 741-748.
7. Holten, D., & Van Wijk, J. J. (2009, April). A user study on visualizing directed edges in graphs. In *Proceedings of the SIGCHI Conference on Human Factors in Computing Systems* (pp. 2299-2308).
8. Holten, D., & Van Wijk, J. J. (2009, June). Force-directed edge bundling for graph visualization. In *Computer Graphics Forum* (Vol. 28, No. 3, pp. 983-990). Oxford, UK: Blackwell Publishing Ltd.
9. Jenny, B., Stephen, D. M., Muehlenhaus, I., Marston, B. E., Sharma, R., Zhang, E., & Jenny, H. (2018). Design principles for origin-destination flow maps. *Cartography and Geographic Information Science*, 45(1), 62-75.
10. Lhuillier, A., Hurter, C., & Telea, A. (2017, June). State of the art in edge and trail bundling techniques. In *Computer Graphics Forum* (Vol. 36, No. 3, pp. 619-645).
11. Rae, A. (2009). From spatial interaction data to spatial interaction information? Geovisualisation and spatial structures of migration from the 2001 UK census. *Computers, Environment and Urban Systems*, 33(3), 161-178.
12. Schläpfer, M., Dong, L., O'Keeffe, K. et al. The universal visitation law of human mobility. *Nature* 593, 522–527 (2021). https://doi.org/10.1038/s41586-021-03480-9
13. Selassie, D., Heller, B., & Heer, J. (2011). Divided edge bundling for directional network data. *IEEE Transactions on Visualization and Computer Graphics*, 17(12), 2354-2363.
14. United Nations. 68% Of the World Population Projected to Live in Urban Areas by 2050, Says UN. Retrieved June 10, 2021 from https://www.un.org/development/desa/en/news/population/2018-revision-of-world-urbanization-prospects.html.
15. Wood, J., Dykes, J., & Slingsby, A. (2010). Visualisation of origins, destinations and flows with OD maps. *The Cartographic Journal*, 47(2), 117-129.
16. Wood, J., Slingsby, A., & Dykes, J. (2011). Visualizing the dynamics of London's bicycle-hire scheme. *Cartographica: The International Journal for Geographic Information and Geovisualization*, 46(4), 239-251.